\documentclass[useAMS,usenatbib]{mn2e}
\usepackage{graphicx}
\usepackage{epsfig}

\title[HD molecules at high redshift]
      {HD molecules at high redshift: \\
       The absorption system at $z=2.3377$ towards Q 1232+082
       \thanks{Based on observations carried out at the ESO
               with the UVES mounted on
               the VLT, on Cerro Paranal in Chile,
               under progs. ID 65.P-0038, 71.B-0136 (P.I. Srianand), and 68.A-0106, 69.A-0061, 70.A-0017 (P.I. Petitjean)}}
\author[A.V. Ivanchik et al.]
{A.V.~Ivanchik$^1$\thanks{E-mail: iav@astro.ioffe.ru (AVI)},
P.~Petitjean$^2$, S.A.~Balashev$^1$, R.~Srianand$^3$,
\newauthor  D.A.~Varshalovich$^1$, C.~Ledoux$^4$, and P.~Noterdaeme$^3$\\
\vspace{7pt}\\
$^1$Ioffe Physical-Technical Institute of RAS, 194021 Saint-Petersburg, Russia\\
$^2$Institut d'Astrophysique de Paris, Universit\'e Paris 06 \& CNRS, UMR7095, 98bis bd Arago, 75014
Paris, France \\
$^3$IUCAA, Post Bag 4, Ganesh Khind, Pune 411~007, India \\
$^4$European Southern Observatory, Alonso de C\'ordova 3107, Casilla
19001, Vitacura, Santiago 19, Chile }

\begin{document}

\date{Accepted 2010 January 19. Received 2010 January 18; in original form 2009 October 28}

\pagerange{\pageref{firstpage}--\pageref{lastpage}} \pubyear{2009}

\maketitle

\label{firstpage}

\begin{abstract}
We present a detailed analysis of the H$_2$ and HD absorption lines
detected in the Damped Lyman-$\alpha$ (DLA) system at  $z_{\rm
abs}=2.3377$ towards the quasar Q~1232+082. We show that this
intervening cloud has a covering factor smaller than unity and
covers only part of the QSO broad emission line region. The zero
flux level has to be corrected at the position of the saturated
H$_2$ and optically thin HD lines by about 10\%. We accurately
determine the Doppler parameter for HD and C~{\sc i} lines
($b$~=~1.86$\pm$0.20~km/s). We find a ratio $N({\rm HD})$/$N({\rm
H_2})$~=~$(7.1^{+3.7}_{-2.2})\times 10^{-5}$ that is significantly
higher than what is observed in molecular clouds of the Galaxy.
Chemical models suggest that in the physical conditions prevailing
in the central part of molecular clouds, deuterium and hydrogen are
mostly in their molecular forms. Assuming this is true, we derive
D/H~$=$~(3.6$^{+1.9}_{-1.1})\times10^{-5}$. This implies that the
corresponding baryon density of the Universe is $\Omega_{\rm b} h^2
= 0.0182^{+0.0047}_{-0.0042}$. This value coincides within 1$\sigma$
with that derived from observations of the CMBR as well as from
observations of the D/H atomic ratio in low-metallicity QSO
absorption line systems. The observation of HD at high redshift is
therefore a promising independent method to constrain $\Omega_{\rm
b}$. This observation indicates as well a low astration factor of
deuterium. This can be interpreted as the consequence of an intense
infall of primordial gas onto the associated galaxy.

\end{abstract}


\section{Introduction}

\begin{figure*}
  \includegraphics[width=175mm,clip]{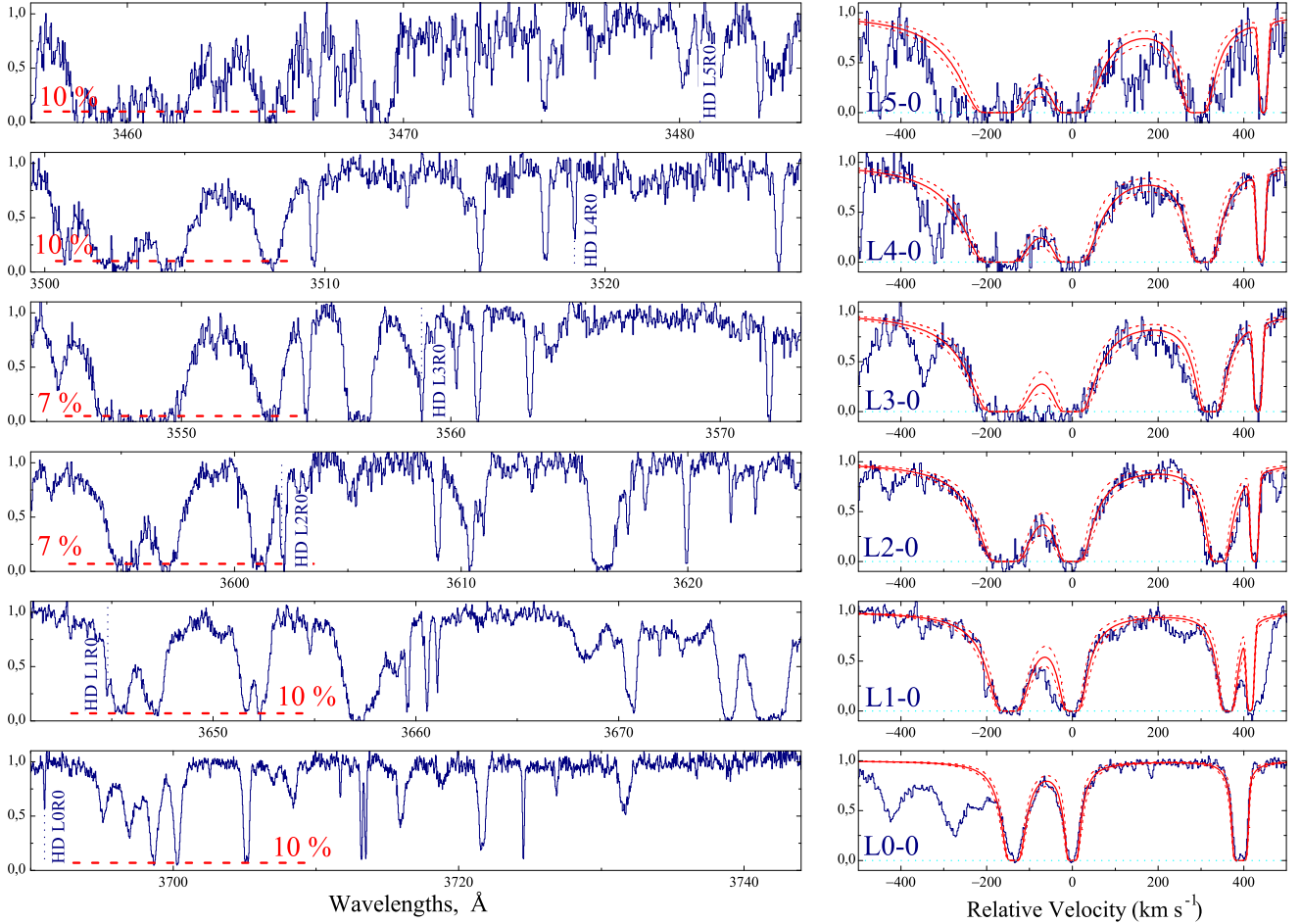}
  \caption{{\sl Left panels}: Portions of the Q~1232+082 spectrum where
           H$_2$ (from L0-0 to L5-0 Lyman bands) and HD lines at $z=2.33771$ are
           redshifted. It can be seen that some H$_2$ (J~=~0 and 1) lines,
           although saturated and showing damped wings, do not
           reach the zero level.
           The dashed horizontal lines show the residual intensity level.
           {\sl Right panels}: Voigt-profile fits of H$_2$ lines once the effect
           of partial covering factor has been taken into account
           (the dotted horizontal lines mark zero level).
           The profiles are plotted on a velocity scale with origin at $z$~=~2.33771.}
  \label{Spectrum}
\end{figure*}

Molecular hydrogen in its two forms, H$_2$ and HD, being the main
cooling agent of primordial gas, plays a central role in the
evolution of gas condensations and the formation of the first
objects in the post-recombination Universe (see \citet{Puy93},
\citet{Palla95}, \citet{Lepp02}, \citet{McGreer08},
\citet{Bromm09}). Observations of these molecules in high-redshift
clouds provide us with clues on the physico-chemical processes at
work and the physical conditions prevailing in the early Universe.

The other important issue associated with deuterium follows from the
fact that, among the light elements created by Primordial
Nucleosynthesis, deuterium is the one whose abundance is most
sensitive to the baryon density of the Universe, $\Omega_{\rm b}$,
or equivalently the baryon-to-photon ratio, $\eta \equiv n_{\rm
b}/n_{\gamma}$, (e.g. \citet{Sarkar96}, \citet{Olive00},
\citet{Coc2005}, \citet{Fields2006}, \citet{Pettini08}). The
primordial deuterium abundance can be directly constrained from
observation of deuterium absorption lines in high-redshift quasar
spectra formed about 10-13 Gyrs ago. So far, all available D/H
measurements at high-redshift are based on the determination of the
column density ratio $N$(D\,{\sc i})/$N$(H\,{\sc i}) in
low-metallicity QSO absorption systems. However, this method
encounters a number of difficulties. Firstly, the velocity
separation between D\,{\sc i} and H\,{\sc i} absorption lines due to
the isotope shift is rather small ($\Delta v \sim -81.6
\,\,$km$\,$s$^{-1}$). This implies that the $N$(H\,{\sc i}) column
density must be small enough so that the D\,{\sc i} lines are not
blended into the corresponding H\,{\sc i} lines but, at the same
time high enough so that the D\,{\sc i} lines can be detected.
Secondly, the presence of the Lyman-$\alpha$ forest makes the
detection of the D\,{\sc i} Lyman-$\alpha$ line somewhat ambiguous
because we cannot be sure that the lines treated as D\,{\sc i}
Lyman-$\alpha$ are not weak H\,{\sc i} Lyman-$\alpha$ from another
intervening cloud.

This explains why, in spite of intensive efforts, only nine
relevant measurements have been performed to date (Pettini et al.
2008 and references therein). Moreover, it should be noted that the
dispersion in the measurements exceeds the individual errors. This
may indicate either the individual errors are underestimated or the
method suffers from systematic effects (see a discussion of the
difficulties inherent to the method by Steigman, 2007).

Here, we discuss an alternative way of constraining the primordial
D/H ratio using the HD/H$_2$ ratio measured in low-metallicity
high-redshift molecular clouds. Such investigations became possible
after the discovery of the first molecular absorbing cloud at high
redshift (Levshakov \& Varshalovich 1985). However, molecules at
high redshift are rarely detected. Indeed, H$_2$ is detected in
absorption in only about 10$\,$\% of the Damped Lyman-$\alpha$ (DLA)
systems and to date only 14 H$_2$ detections have been reported
(Ledoux et al. 2003, Noterdaeme et al. 2008a). HD is seen in only
two of these systems (see \citet{Varsh2001} for the first one and
\citet{Srian2008} for the second one). The second HD absorption
system towards SDSS~J143912+111740 was studied in detail by
\citet{Noterd2008b}. They found $N_{\rm HD}$/$2N_{\rm
H_2}$~=~$(1.5^{+0.6}_{-0.4})\times 10^{-5}$ for a metallicity close
to solar and a molecular fraction
$f$~=~2$N$(H$_2$)/[2$N$(H$_2$)~+~$N$(H$\,${\sc
i})]~=~0.27$^{+0.10}_{-0.08}$. This is significantly higher than
what is observed in the ISM of the Galaxy for a comparable
metallicity ($0.4\div4.3\times10^{-6}$, Lacour et al., 2005a). This
can be understood if a large fraction of the deuterium in the gas
originates from pristine gas infalling onto the host galaxy. Here,
we present a detailed analysis of the first observed HD absorption
system at $z=2.3377$ towards QSO~1232+082. We present the
observations in Section~2, the data analysis and the observed
HD/H$_2$ and D/H ratios in Section~3 before concluding in Section~4.

\section{Observations}
The high spectral resolution spectrum of the high redshift quasar
PKS~1232+082 ($z_{\rm em}=2.57$ and $m_{\rm V}=18.4$) was obtained
over several observing campaigns in the course of a survey for and
follow-up studies of molecular hydrogen in DLA systems with the
Ultraviolet and Visible Echelle Spectrograph (UVES) mounted on the
ESO Kueyen VLT telescope on Cerro Paranal in Chile (Petitjean et al.
2000, Ledoux et al. 2003, Noterdaeme et al. 2008a). We used Dichroic
\#1 and central wavelengths 390 and 564~nm in the blue and red
spectroscopic arms respectively. The total exposure time on the
source was 17.5$\,$h. The CCD pixels were binned 2$\times$2 and the
slit width adjusted to 1'' matching the mean seeing conditions
of$\,\sim\,$0.9''. This yields a resolving power of
R$\,\sim\,$45,000. The data were reduced using the UVES pipeline
based on the ESO MIDAS system. Wavelengths were rebinned to the
vacuum-heliocentric rest frame and individual scientific exposures
co-added using a sliding window and weighting the signal by the
total errors in each pixel.

The observed QSO spectrum contains the log~$N$(H~{\sc i})~=~20.90$\pm$0.08
DLA H$_2$ absorption system at $z_{\rm abs}=2.33771$ that was originally
detected by \citet{Ge99} and analyzed by Srianand et al. (2000), and in which
HD molecules were seen for the first time by \citet{Varsh2001}.

\section{Analysis of the absorption system}
The detailed analysis of this absorption system reveals two
interesting peculiarities: (i) firstly, as it is apparent in
Fig.~\ref{Spectrum} and Fig.~\ref{Level10}, at least the central
part of the absorbing cloud does not cover the background source
completely (the covering factor of the cloud is not unity) and (ii)
secondly, the observed Doppler parameter, $b$, is larger for H$_2$
lines ($\sim$4.5 km/s) than for HD and C$\,${\sc i} lines ($\sim$1.9
km/s). These two effects (described in Sec. 3.1 and 3.2) have
important consequences on the determination of the column densities
and HD/H$_2$ ratio (Sec. 3.3 and 3.4).

\subsection{Covering factor effect}

\begin{figure}
\includegraphics[width=83mm,clip]{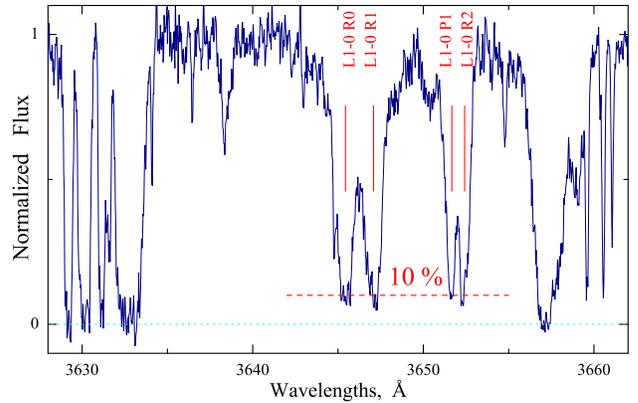}
 \caption{Portion of the Q~1232+082 spectrum showing that the profiles of strong
          saturated H$_2$ absorption lines (L1-0 R0, R1, P1, and R2) redshifted
          ($z_{\rm abs}$~=~2.33771) on top of the QSO Lyman-$\beta$ and O$\,${\sc vi}
          emission lines ($z_{\rm em}=2.57$)
          do not go to the zero flux, while the strong intervening Lyman-$\alpha$
          lines do go to the zero flux with an mean uncertainty of $\sim2$~\%.
}
 \label{Level10}
\end{figure}

\begin{figure}
\includegraphics[width=83mm,clip]{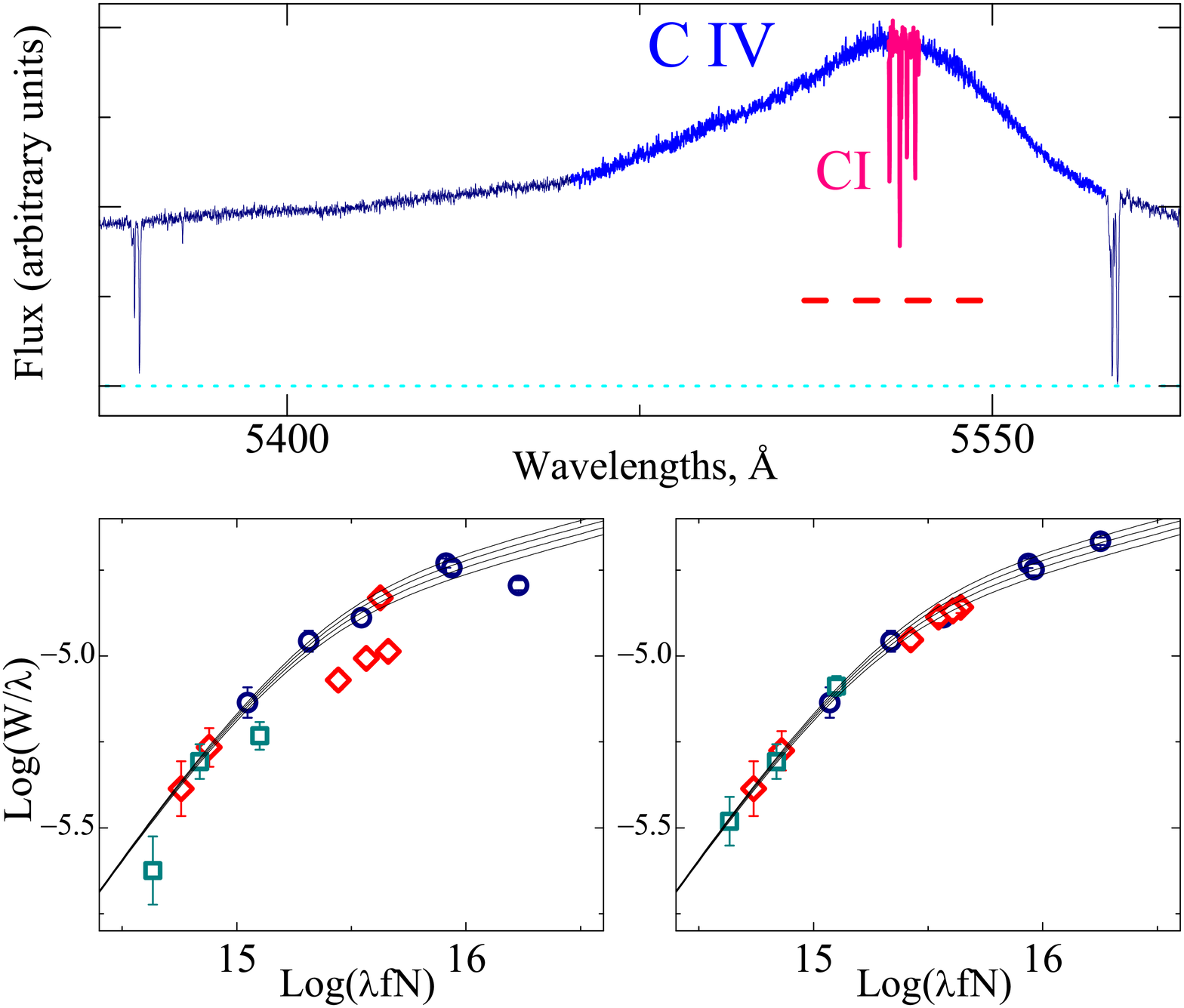}
 \caption{{\sl Upper panel}: Portion of the QSO~1232+082 spectrum showing the
six C$\,${\sc i} absorption lines associated with the H$_2$ absorption
system ($z_{\rm abs}$~=~2.33771) and redshifted on top of the
C$\,${\sc iv} QSO emission line ($z_{\rm em}=2.57$). {\sl Lower
panels}: The positions of the 15 C$\,${\sc i} absorption lines
detected in the spectrum are shown in the standard curve of growth
diagram (left hand side panel). Lines from the true ground-state,
the first and second fine structure excited levels are indicated in,
respectively, dark blue, red and green. The six lines that are
located on top of the C$\,${\sc iv} emission line are systematically
shifted downwards relative to the curve of growth followed by the
other lines (black solid lines; for those lines the different column densities
are derived from optically thin lines redshifted
away from QSO emission lines). Assuming that the six
C$\,${\sc i} lines cover only $\sim$75~\% of the QSO emission region
and correcting the equivalent widths from this, brings the absorption
lines back to the curve of growth (right hand side panel).}
 \label{CI}
\end{figure}

It can be seen on Fig.~\ref{Spectrum} that the profiles of strong
saturated H$_2$ absorption lines do not go to the zero flux. Some of
them even show damped wings. This is apparent for the lines
redshifted on top of the QSO Lyman-$\beta$ and O~{\sc vi} emission
lines, in particular it is the case for the L0-0 and L1-0 H$_2$
absorption lines. We have carefully checked that this cannot be due
to an error in the positioning of the zero flux. Indeed, it can be
seen on Fig.~\ref{Level10} that all the strong intervening
Lyman-$\alpha$ lines do go to the zero flux with a mean uncertainty
of $\sim 2$~\% (due to, for instance, uncertainties in the scattered
light subtraction).

This means that part of the QSO radiation is not intercepted by the
molecular cloud. There are several explanations for this, including:
(i) the emission region of the background source has an angular size
larger than the angular size of the absorbing cloud, (ii) there are
two distinct background sources (tight double quasars, e.g. Foreman
et al. 2009) or (iii) the quasar is lensed which implies the
presence of several unresolved images.

To our knowledge this is the first time such an intervening cloud is
observed and we believe this deserves a detailed discussion that we
postpone to a forthcoming paper (Balashev et al. 2009a). Here, we
will correct the column density determinations from this
partial covering factor and concentrate on the HD content of the cloud.

We can ascertain the partial covering factor of the cloud using the
numerous C$\,${\sc i} absorption lines that are detected in the
spectrum. In particular six lines are located on top of the
C$\,${\sc iv} emission line (see Fig.~\ref{CI}). One line originates
from the true C$\,${\sc i} ground state whereas, respectively, three
and two lines originate from the first and second fine structure
excited states (C$\,${\sc i}$^*$ and C$\,${\sc i}$^{**}$ absorption
lines). We plot the positions of these lines together with other
C$\,${\sc i} lines observed in the spectrum in a standard curve of
growth diagram giving log~($W/\lambda$) versus log~($\lambda fN$)
(see left hand side lower panel of Fig.~\ref{CI}). It can be seen
that the six absorption lines located on top of the C$\,${\sc iv}
emission line are situated below the curve of growth followed by the
other lines. We have to correct for a covering factor of $\sim$75~\%
to recover a unique curve of growth (right hand side lower panel of
Fig.~\ref{CI}). This implies that the cloud covers the QSO continuum
emission region completely (ascertained by the fact that e.g. the
C$\,${\sc ii}$\lambda$1334 line is going to zero) but only part of
the broad line region. The covering factor depends on the considered
emission line because regions emitting different emission lines have
different sizes.

In the following, we aim at deriving the column densities of H$_2$
and HD. Since the H$_2$ lines are saturated, it is easy to correct
for the partial covering factor by correcting the zero level.
However the HD lines are not strongly saturated and we need to
estimate the covering factor of these lines. Three HD absorption
lines L5-0\,R0, L4-0\,R0 and L-00\,R0 are unblended and will be used
to derive $N$(HD). It can be seen on Fig.~1 that L5-0 is redshifted
at $\sim$\,3480~\AA, on top of the QSO Lyman-$\gamma$ emission line,
L4-0 is redshifted at $\sim$\,3520~\AA, on top of C$\,${\sc
iii}$\lambda$987 QSO emission line, and L0-0 is redshifted at
$\sim$3690~\AA, on top of the O$\,${\sc vi}$\lambda\lambda$1031,1037
QSO emission line (see composite quasar spectra in e.g. Vanden Berk
et al. 2001). As the H$_2$ absorption lines in these wavelength
ranges all have a covering factor of $\sim$90\% and since most of
the H$_2$ must be co-spatial with HD, we apply the same covering
factor to the HD absorption lines (see Section~\ref{CD}).

\subsection{Broadening effect}
\label{BE}

Whereas an accurate determination of $N^{\rm tot}_{{\rm H}_2}$ is
possible because the most important H$_2$ transitions are damped,
the determination of the column densities for moderately saturated
lines like the HD lines strongly depends on the Doppler parameter.
This is why it is important to study the broadening of the lines in
details.

\begin{figure}
\includegraphics[width=83mm,clip]{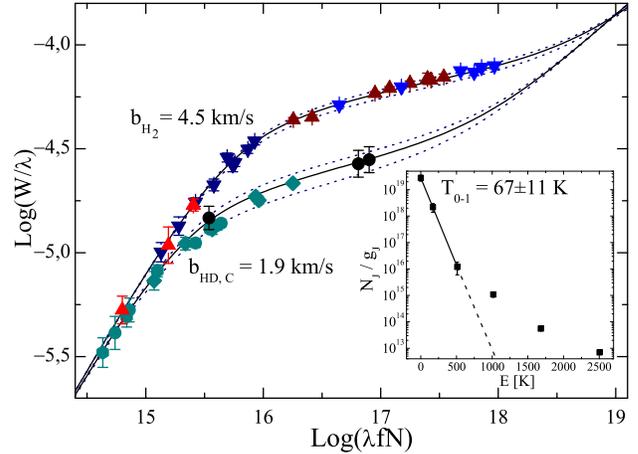}
 \caption{Curves of growth for $H_2$ (indicated in red, dark-blue, brown,
          and blue triangles, respectively,  for J\,=\,5,\,4,\,3, and 2 levels),
          HD (black circles), and C$\,${\sc i} (cyan circles) absorption lines at
          $z_{\rm abs}$~=~2.33771 toward QSO~1232+082. {\sl Inset}:
          H$_2$ excitation diagram from which we measure
          $T_{0\!-\!1}$~=~$T_{\rm kin}$~=~67$\pm11$~K.
          }
 \label{All_CG}
\end{figure}

We show in Fig.~\ref{All_CG} the curves of growth of the H$_2$,
C$\,${\sc i} and HD lines after correction for partial covering
factor. It is apparent that a unique Doppler parameter describes the
curve of growth of all the H$_2$ lines on the one hand and of the
C$\,${\sc i} and HD lines on the other. Corresponding values are
$b_{\rm H2}$~$\sim$~4.5~km~s$^{-1}$ and $b_{\rm CI} \simeq b_{\rm
HD} \sim $~1.9~km/s. The corresponding column densities are given in
Table~1. The Doppler parameter is observed to be the same for all
H$_2$ rotational levels with a value much larger than what is
expected from thermal broadening. Indeed, the excitation diagram is
shown as an inset in Fig.~\ref{All_CG} and the excitation
temperature is found to be \mbox{T$_{0\!-\!1}=67\pm11 \,$K} for the
three lower rotational levels and larger for higher excitation
levels. The first temperature is thought to reflect the kinetic
temperature of the gas and corresponds to a thermal component of the
Doppler parameter of $b_{\rm th}^{\rm CI,HD}<b_{\rm th}^{\rm H2}<0.8
\,$km/s. This means that the observed Doppler parameters are mostly
due to turbulent motions in the gas.

It has been observed in a few cases that the Doppler parameter of
H$_2$  lines increases with the rotational level (see
\citet{Spit73}; \citet{Jenkins97}; \citet{Lacour05b} for
observations in the Galaxy and \citet{Noterd2007} for observations
in a DLA at high redshift). \citet{B2009b} have shown that this can
be interpreted as a consequence of (i) radiative transfer effect:
the directional radiation can lead to broadening of the high
rotation level velocity distribution due to saturation of radiative
pumping lines; (ii) the structure of the cloud: absorption from high
rotational levels arise mostly from the external part of the cloud
where temperature and turbulence are larger. In addition, not only
C$\,${\sc i} atoms are ionized more easily than H$_2$ is destroyed,
but H$_2$ is more effectively self-shielded than HD. Therefore
C$\,${\sc i} and HD species are expected to be located in the
central part of the cloud where turbulence is minimal and H$_2$
molecules are mostly in their low rotational levels. Since we
determine the H$_2$ $b$ value from high rotational levels, it is not
surprising to derive different a larger Doppler parameter for H$_2$
compared to C$\,${\sc i}--HD.

\begin{table*}
 \centering
 \begin{minipage}{157mm}
  \caption{Column densities and Doppler parameters measured for H$_2$ and HD molecules.}
    \label{Exp_Data}
  \begin{tabular}{@{}cccccc@{}}
  \hline\hline
    & H$_2$ &  &  & HD & \\
  \hline
    ~Rotational & log~$N$(J)   & $b$           & ~Rotational & log~$N$(J)  & ~$b$ ~ \\
     Level      &  (cm$^{-2}$) & (km s$^{-1}$) & Level      & (cm$^{-2}$) & (km s$^{-1}$)\\
  \hline
  J=0& 19.45$\pm$0.10 &       & J=0~ & 15.53$^{+0.17}_{-0.11}$  & 1.86$\pm$0.20   \\
  J=1& 19.29$\pm$0.15 &       &  & (15.51$\pm$0.13)$^\star$ & (1.58$\pm$0.18)$^\star$  \\
  J=2& 16.78$\pm$0.24 &  4.5  & J=1~ & $<14.0$ & \\
  J=3& 16.36$\pm$0.10 &  4.5  &&& \\
  J=4& 14.70$\pm$0.06 & $>3.3$&&& \\
  J=5& 14.36$\pm$0.07 &       &&& \\
\hline
  & $N^{\rm tot}_{{\rm H}_2}\!=\!(4.78^{+0.96}_{-0.96})\!\times\!10^{19}\,$cm$^{-2}$ & &
  & $N^{\rm tot}_{\rm HD}\!=\!(3.39^{+1.6}_{-0.8})\!\times\!10^{15}\,$cm$^{-2}$ &  \\
\hline
  \multicolumn{6}{c}{$N_{\rm HD}/N_{{\rm H}_2}\!=\!(7.1^{+3.7}_{-2.2})\!\times\!10^{-5}$}\\
\hline
\end{tabular}

{\footnotesize $^\star$values calculated without taking into account
covering factor}

\end{minipage}
\end{table*}

\subsection{H$_2$ and HD column densities}
\label{CD}

\begin{figure}
\includegraphics[width=83mm,clip]{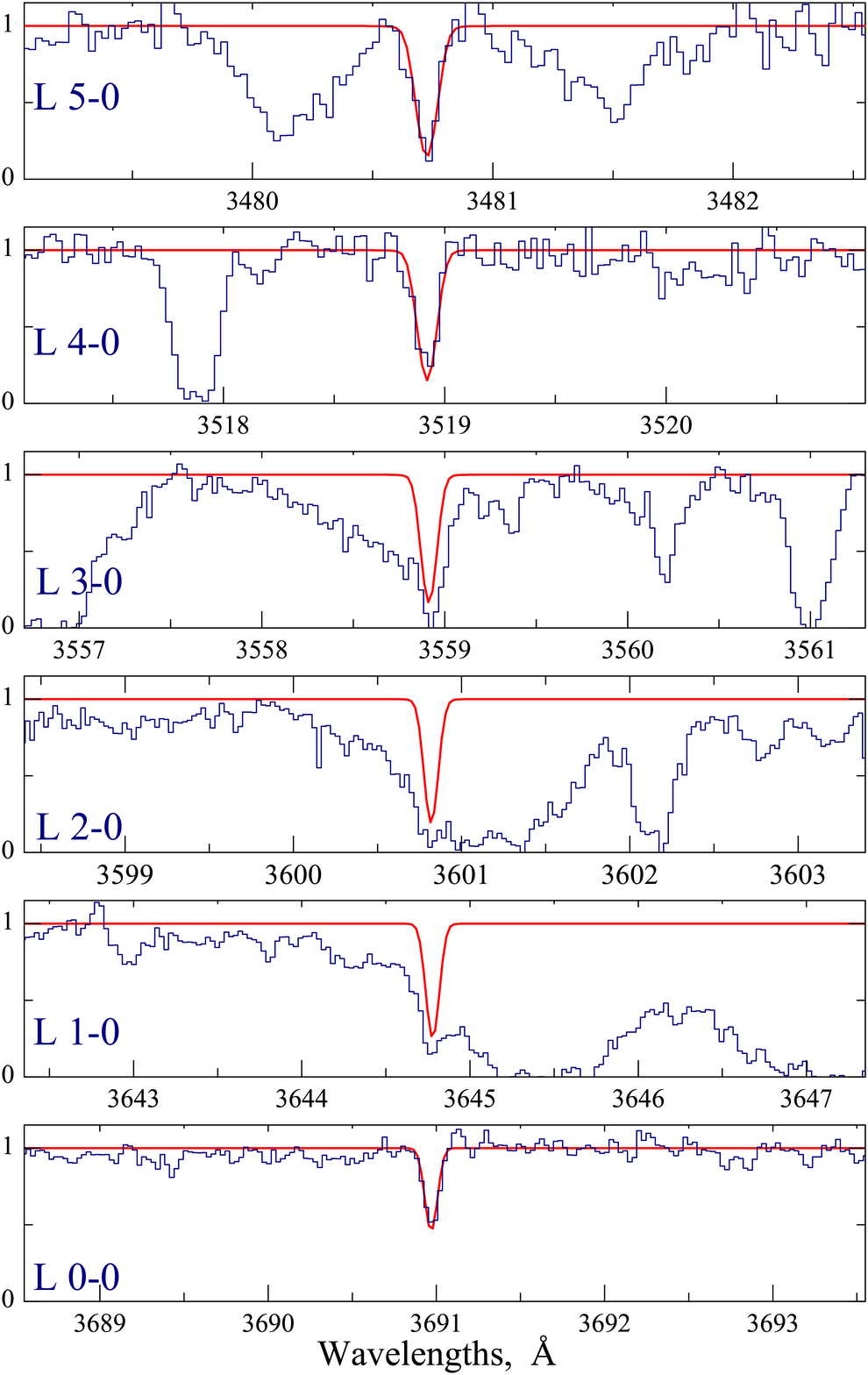}
 \caption{Voigt-profile fits to the six HD Lyman band absorption lines from
          the J~=~0 rotational level.
          Three lines (L0R0, L4R0, L5R0) are well defined and unblended
          and can be used in the analysis while the other three lines are detected
          but strongly blended and are not used to constrain the HD column density.}
 \label{Spec_HD}
\end{figure}

Column densities of H$_2$ in different rotational levels (J~=~0 to
5) are  obtained by simultaneous fits to the Lyman 0-0 to 5-0 band
transition lines (see Fig.~\ref{Spectrum}). The results of the fits
are presented in Table~1. The total H$_2$ column density is
\mbox{$N^{\rm tot}_{{\rm
H}_2}\!=\!(4.78\pm0.96)\!\times\!10^{19}\,$cm$^{-2}$}. This is
consistent within 1$\sigma$ with the results by Noterdaeme et al.
(2008a). Note that the smaller value obtained by Srianand et al.
(2000) is due to the fact that partial covering factor of L0-0 lines
was not recognized because of the much lower signal-to-noise ratio
of the data available at that time. This implies a molecular
fraction of
log~$f$~=~log[2$N$(H2)/($N$(HI)+2$N$(H2))]~=~$-$0.92$\pm$0.15 if we
assume that the neutral hydrogen is co-spatial with the molecules.
We note that this is therefore probably a lower limit of the
molecular fraction in the molecular part of the cloud.

The first rotational level of deuterated molecular hydrogen is
detected in six (from 0-0 to 5-0) Lyman bands covered by the
observations (Fig.~\ref{Spec_HD}). Unfortunately, three of six lines
are strongly blended and cannot be used to derive line parameters.
Therefore we use only the L5-0, L4-0 and L0-0 transitions.
Laboratory wavelengths and oscillator strengths for the three lines
are given in Table~2. We apply a 10\% correction factor to the zero
flux at the position of the lines equal to what is observed for the
H$_2$ lines located close to their position in the spectrum (see
Fig.~\ref{Spectrum}). To convince ourselves that such covering
factor must be applied, we note that HD-L4-0 at $\sim$3519~\AA~ is
located only 1~\AA~ away from a saturated H$_2$ line at
$\sim$3519~\AA~ that has to be corrected from this covering factor.
However, we have checked that if we do not correct for partial
covering factor, the best fit of the lines yields very similar
result (log~$N$(HD)~$\sim$~15.51 and $b$~$\sim$~1.58~km~s$^{-1}$;
see Table~1). The following discussion is therefore done in the case
of the lines being corrected.

\begin{figure}
\includegraphics[width=83mm,clip]{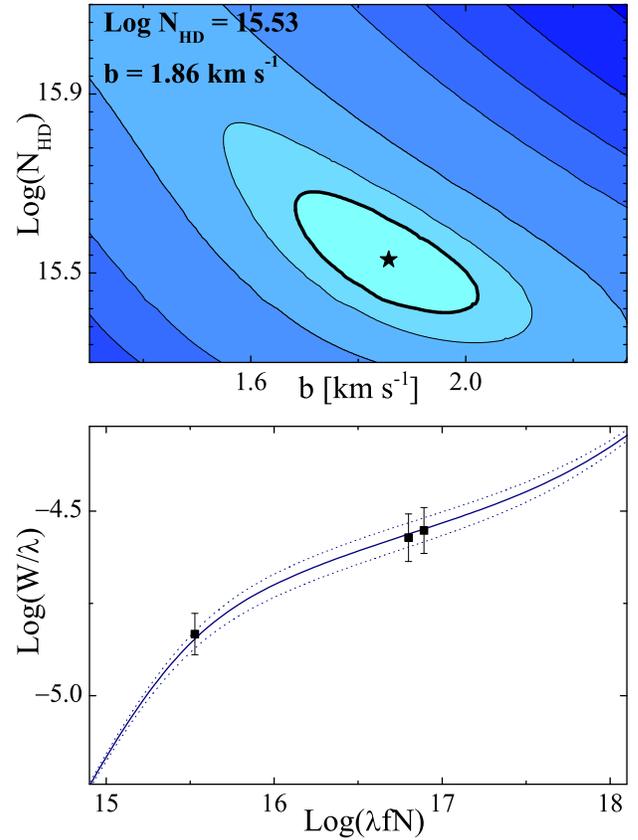}
 \caption{Results of a $\chi^2$ minimization analysis of the three unblended
          HD lines (L0R0, L4R0, L5R0, see Fig.~\ref{Spec_HD}) yielding the HD
          column density and $b$-value.
          {\sl Top panel}: Best model and confidence regions (the thick black line is 1$\sigma$ contour).
          {\sl Bottom panel}: Corresponding curve of growth for
          $b$~=~1.86~km$\,$s$^{-1}$ and positions
          of the HD lines for log~$N_{\rm HD}$(cm$^{-2}$)~=~15.53.}
 \label{LogN_b}
\end{figure}

The fit of the three lines is degenerate and leads to two equally
acceptable solutions, consistent with two different positions of
the curve of growth: either
$b$~=~0.89~km/s and log~$N$(HD)~=~16.55 or $b$~=~1.86~km/s and
log~$N$(HD)~=~15.52. Since the strength of the 15 observed C~{\sc i}
lines are consistent with $b$~=~1.86~km/s (see Fig.~\ref{All_CG}),
we favor the second solution. Indeed we expect C$\,${\sc i} and HD
to be co-spatial (see discussion at the end of section~\ref{BE}).
Results of analysis are illustrated in Fig.~\ref{LogN_b}.

The total HD column density is therefore \mbox{$N^{\rm tot}_{\rm
HD}\!=\!(3.39^{+1.6}_{-0.8})\!\times\!10^{15}\,$cm$^{-2}$} (note
that the calculated $N^{\rm tot}_{\rm HD}$ error includes the error
on the Doppler parameter, b) and the ratio $N_{\rm HD}$/$N_{\rm
H_2}$=$(7.1^{+3.7}_{-2.2})\times 10^{-5}$ is the largest one ever
observed in any astrophysical objects of the Galaxy
($<8.6\cdot10^{-6}$, Lacour et al., 2005a) and beyond
($3.0\cdot10^{-5}$, Noterdaeme et al., 2008b).

\begin{table}
 \centering
 \begin{minipage}{83mm}

  \caption{Wavelengths and oscillator strengths for HD lines.}

  \begin{tabular}{@{}ccl@{}}
  \hline\hline
    ~~Lines & $\lambda$$^\star$ (\AA) & ~~~~$f$ \\
  \hline
    ~~L0R0  & ~~~~~~~1105.84055~~~~~~~   & 0.00082~ \\
    ~~L1R0  & 1092.00126                 & 0.00311  \\
    ~~L2R0  & 1078.83104                 & 0.0068   \\
    ~~L3R0  & 1066.27568                 & 0.0114   \\
    ~~L4R0  & 1054.29354                 & 0.0161   \\
    ~~L5R0  & 1042.85005                 & 0.0201   \\

  \hline

\end{tabular}

{\footnotesize $^\star$HD laboratory wavelengths from
\mbox{\citet{Iv2008}} }

\end{minipage}
\end{table}

\subsection{The {\bf D/H} ratio from HD/H$_2$}

The absorption system at $z_{\rm abs}$~=~2.3377 toward Q~1232+082
has one of the lowest total metallicities among the 14 H$_2$ system
known to date, [S/H]~=~$-1.43\pm0.08$, for one of the highest
molecular fraction (log~$f$~=~$-0.92\pm0.15$, see also Noterdaeme et
al. 2008a). This implies that at least the densest and most central
parts of the cloud has a chemical composition close to primordial.
We can therefore use the HD/H$_2$ ratio to estimate the primordial
D/H ratio.

Certainly, the chemistry of molecular clouds makes the determination
of the D/H isotope ratio from HD/H$_2$ rather uncertain.
Nevertheless, molecular cloud models (e.g. Le~Petit et al., 2002,
that is an extension of a previous PDR model, Abgrall et al., 1992;
Le Bourlot, 2000) suggest that deuterium is molecular only when
hydrogen is predominantly in its molecular form. In the case of
Q~1232+082, we are probably very close to the conditions at which
the D/HD transitions occurs. Indeed, this transition happens for
$n_{\rm H} \sim 300 \,$cm$^{-3}$, $T \sim 60 \,$K, $A_{\rm V} \sim
0.3$ (when the fractions of H in H$_2$ and D in HD are less than
4\%) that corresponds approximately to what is measured in the DLA
(Srianand et al. 2000, 2005; Noterdaeme et al. 2008). The formation
of HD in that case is dominated by the D$^+$+H$_2$ interaction (Le
Petit et al. 2002). FUSE and Copernicus observations have also shown
that in the ISM of the Galaxy, the HD/H$_2$  ratio increases with
the molecular fraction and that HD/2H$_2$ could trace D/H well when
$f \sim 1$ (Lacour et al. 2005a; Linsky et al., 2006; Snow et al.,
2008), when both HD and H$_2$  are self-shielded from
photo-dissociation. Self-shielding becomes efficient as soon as the
optical depth $\tau$ of the discrete absorbing transitions is of the
order of 1, that is when the column densities of HD and H$_2$ are
about $10^{14}\!\div10^{15}$cm$^{-2}$. In our case, they are
$3.39\cdot10^{15}$ and $4.78\cdot10^{19}$cm$^{-2}$, respectively.
All this supports the idea that the conditions in the cloud are
close to what is needed to have D/H $\sim$ HD/2H$_2$ within our
measurement errors.

Under the above assumption we thus derive
D/H~$\sim$~HD/2H$_2$\,$\,\sim\,$\,(3.6$^{+1.9}_{-1.1}$)$\times$10$^{-5}$
in the cloud. This corresponds to a negligible astration factor.
Since the total metallicity in the absorbing cloud is not pristine,
this implies that the associated object has probably undergone
strong infall of primordial gas as already noted by Noterdaeme et
al. (2008a).

\section{Constraining $\Omega_{\rm b}$}

\begin{figure}
\includegraphics[width=83mm,clip]{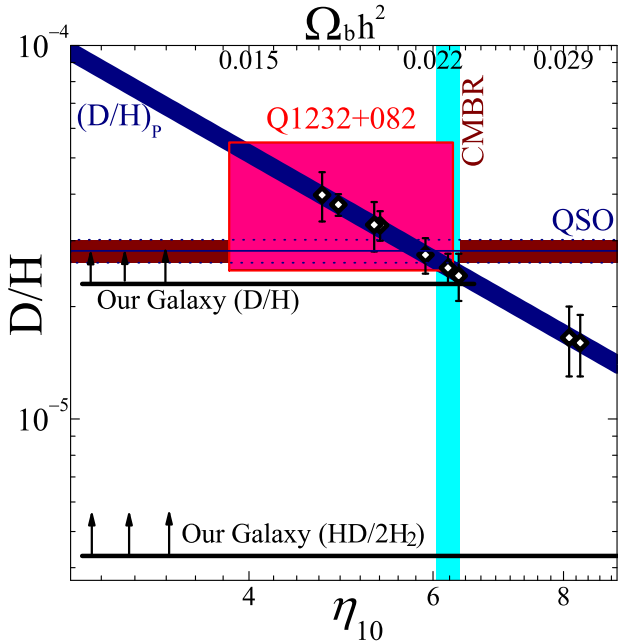}
 \caption{Constraints on the baryon-to-photon ratio $\eta$ (scale at the bottom)
          or equivalently on the baryon density $\Omega_{\rm b}$ (scale at the top)
          from different observations (for numerical values and references
          see Table~\ref{Exp_D_H}). The results of primordial nucleosynthesis are
          indicated as an inclined violet line and our result is shown as the red error box.
          It is still in agreement with the $\Omega_{\rm b}$ estimate from CMBR observations
          (cyan vertical strip) and the D/H isotope measurements in low
          metallicity clouds toward high-redshift quasars (marked as 'QSO' brown
          horizontal strip). However, we note that the scatter of the individual
          points from measurement of D/H in low metallicity clouds (points
          with error bars) is quite large and in any case
          larger than the final error quoted. Therefore the HD/H2 technique could
          lead to interesting results when very high SNR data will be obtained from
          Extremely Large Telescope.
}
 \label{Pr_D}
\end{figure}

Constraints on the baryon-to-photon ratio $\eta$ or equivalently on
the baryon density $\Omega_{\rm b}$ from different observations are
summarized in Table~\ref{Exp_D_H} and illustrated in
Fig.~\ref{Pr_D}. The results of primordial nucleosynthesis are
indicated as an inclined violet line and our result is shown as the
red error box. The $\Omega_{\rm b}$ value derived from CMBR analysis
is indicated by a vertical cyan strip. Pettini et al. (2008) mean
value obtained from a compilation of D$\,${\sc i}/H$\,${\sc i}
measurements in low-metallicity high-redshift clouds is indicated as
an horizontal brown strip. However, Pettini et al. (2008) do not
consider the Levshakov et al. (2002), Crighton et al. (2004) and
Pettini \& Bowen (2001) measurements when calculating this mean
value. Indeed, nine determinations of the D/H ratio in
low-metallicity high-redshift clouds are reported in the literature.
They are shown as open circles with error bars. Whatever points are
used, it is apparent from Fig.~\ref{Pr_D} that the scatter in the
D~{\sc i}/H~{\sc i} measurements is much larger than the errors
quoted by Pettini et al. (2008). This shows that uncertainties from
all methods are large and that the derivation of D/H from
measurements of HD/H$_2$ could nicely complement other measurements
especially with the advent of future large telescopes.

\begin{table*}
 \centering
 \begin{minipage}{147mm}
  \caption{Results from different experiments on D/H ratio determination.}
  \label{Exp_D_H}
  \begin{tabular}{@{}lcccc@{}}
  \hline\hline
    ~~~~~~~~            & ~~Redshift~~~ & D/H & $\Omega_{\rm b} h^2$ & ~Ref.$^a$~ \\
  \hline

  ~~Local Galactic Disk & {\bf 0} & ${\bf > 2.31\times10^{-5}}$ & &  1 \\
\vspace{2mm}
  ~~HD/2H$_2$           & {\bf 0} & ${\bf > 3.7\times 10^{-7} \div 4.3\times 10^{-6}}$ & & 2\\
  ~~~Q2206$-$199       & 2.08 & $(1.65^{+0.35}_{-0.35})\times10^{-5}$ & & 3 \\
  ~~$^*$Q1009+2956      & 2.50 & $(3.98^{+0.59}_{-0.67})\times10^{-5}$ & & 4 \\
  ~~$^*$Q1243+3047      & 2.53 & $(2.42^{+0.35}_{-0.25})\times10^{-5}$ & & 5 \\
  ~~$^*$HS0105+1619     & 2.54 & $(2.54^{+0.23}_{-0.23})\times10^{-5}$ & & 6 \\
  ~~$^*$Q0913+072       & 2.62 & $(2.75^{+0.27}_{-0.24})\times10^{-5}$ & & 7 \\
  ~~$^*$SDSS1558$-$0031   & 2.70 & $(3.31^{+0.49}_{-0.43})\times10^{-5}$ & & 8 \\
  ~~~Q0347$-$3819         & 3.03 & $(3.75^{+0.25}_{-0.25})\times10^{-5}$ & & 9 \\
  ~~~Q1937$-$1009         & 3.26 & $(1.60^{+0.25}_{-0.30})\times10^{-5}$ & & 10 \\
  ~~$^*$Q1937$-$1009      & 3.57 & $(3.30^{+0.30}_{-0.30})\times10^{-5}$ & & 11 \\
  \cline{1-1}
\vspace{2mm}
  ~~$^*$Average         & {\bf 2$-$3.6} & ${\bf (2.82\pm0.20)\times10^{-5}}$  & ${\bf 0.0213\pm0.0010}$ & 7 \\
  ~~CMBR                & ~~{\bf 1500}~~ & & ${\bf 0.02267^{+0.00058}_{-0.00059}}$ & 12 \\
\hline
  ~~HD/2H$_2$           & {\bf 2.3377} & ${\bf (3.6^{+1.9}_{-1.1})\times10^{-5}}$ & ${\bf 0.0182^{+0.0047}_{-0.0042}}$ & ~~this work~~\\
\hline\hline

\end{tabular}

{\footnotesize $^a$References -- (1)~\citet{Linsky06},
(2)~\citet{Lacour05a}, (3)~\citet{Pettini01} (4)~\citet{Burles98b},
(5)~\citet{Kirk03}, (6)~\citet{Omeara01} (7)~\citet{Pettini08},
(8)~\citet{Omeara06}, (9)~\citet{LeV2002}, (10)~\citet{Crig2004},
(11)~\citet{Burles98a}, (12)~\citet{komatsu09} \\ $^*$ --
Measurements used by \citet{Pettini08} to estimate the primordial
D/H ratio. }

\end{minipage}
\end{table*}

\section{Conclusion}

We have analyzed in detail the absorption system at $z_{\rm
abs}$~=~2.3377  towards Q~1232+082 in which H$_2$ and HD molecules
are observed. This is the system where HD detection was reported for
the first time (Varshalovich et al. 2001) and the second system in
which the HD/H$_2$ ratio is studied. We find:

\begin{enumerate}
  \item The intervening absorbing cloud does not cover the background
  source completely. The continuum source is covered but not the totality of
        the broad line region.

  \item The Doppler parameter measured for H$_2$ absorption lines is
  approximately the same for all high rotational levels, $b_{\rm H2}$~=~4.5~km/s,
  and is larger than what is observed for C~{\sc i} and HD transitions,
  $b_{\rm HD, CI}$~=~1.9~km/s.
  The excitation temperature of the three first rotational levels is
  indicative of a kinetic temperature of $\sim$70~K; the excitation
  temperature of the J~=~3 to 5 levels being larger.

  \item We correct for the partial covering effect and measure the H$_2$
  and HD column densities. The total H$_2$ and HD column densities are
  \mbox{$N^{\rm tot}_{{\rm H}_2}\!=\!(4.78\pm0.96)\!\times\!10^{19}\,$cm$^{-2}$}
  and \mbox{$N^{\rm tot}_{\rm HD}\!=\!(3.39^{+1.6}_{-0.8})\!\times\!10^{15}\,$cm$^{-2}$}
  and therefore their ratio $N_{\rm HD}$/$N_{\rm H_2}$~=~$(7.1^{+3.7}_{-2.2})\times 10^{-5}$.
  This is the largest value for HD/H$_2$ ever observed for any astrophysical objects
  in the Galaxy and beyond.
  The HD/H$_2$ ratio in this low-metallicity cloud is significantly larger
  than what is observed in interstellar clouds of the Galaxy.
  The astration factor of deuterium is basically zero.

  \item The physical conditions in the cloud makes it plausible that all hydrogen and
  deuterium are in molecular form. Thus, we conjecture that D/H~$\sim$~HD/2H$_2$
  and find D/H\,$\sim$\,(3.6$^{+1.9}_{-1.1}$)$\times$10$^{-5}$, that
  is very close to the primordial values derived from other techniques.
  We note however that the dispersion of the D/H measurements in clouds with
  primordial abundances is much larger than the given errors on the measurements casting
  some doubts on the error box of the final result.
  Finally it must be reminded that the astrophysical measurements of primordial abundances of
  Li and $^3$He are still not completely consistent with the Big-Bang
  nucleosynthesis (e.g. \citet{Coc2005}, \citet{Cyburt08}). All this
  argues in favor of using the HD/H$_2$ method with better data in order
to better constrain D/H. This
  will be possible with the advent of the foreseen Extremely Large
  Telescopes.

\end{enumerate}

\vspace{7mm}{\footnotesize {\rm Acknowledgments.} This work has been
supported by a bilateral program of the Direction des Relations
Internationales of CNRS in France, by the Russian Foundation for
Basic Research (grant 08-02-01246a), and by a State Program
``Leading Scientific Schools of Russian Federation'' (grant
NSh-2600.2008.2).}


\label{lastpage}


\begin{thebibliography}{99}

\bibitem[\protect\citeauthoryear{Abgrall et al.}{1992}]{A1992}
Abgrall H., Le Bourlot J., Pineau des For\^{e}ts G., 1992, A\&A,
253, 525

\bibitem[\protect\citeauthoryear{Balashev et al.}{2009a}]{B2009a}
Balashev S., Varshalovich D., Ivanchik A., and Petitjean~P., 2009a,
in preparation

\bibitem[\protect\citeauthoryear{Balashev et al.}{2009b}]{B2009b}
Balashev S., Varshalovich D., Ivanchik A., 2009b, Astronomy
Letters, 35, 150

\bibitem[\protect\citeauthoryear{Ge \& Bechtold}{1999}]{Ge99}
Ge J., Bechtold J., 1999, Highly Redshifted Radio Lines, ASP Conf.
Series, 156, 121

\bibitem[\protect\citeauthoryear{Bromm et al.}{2009}]{Bromm09}
Bromm V., Yoshida N., Hernquist L., McKee C.F., 2009, Nature, 459,
49

\bibitem[\protect\citeauthoryear{Burles \& Tytler}{1998a}]{Burles98a}
Burles S., Tytler D., 1998a, ApJ, 499, 699

\bibitem[\protect\citeauthoryear{Burles \& Tytler}{1998b}]{Burles98b}
Burles S., Tytler D., 1998b, ApJ, 507, 732

\bibitem[\protect\citeauthoryear{Coc et al.}{2005}]{Coc2005}
Coc A., Angulo C., Vangioni-Flam E., Descouvemont P., Adahchour A.,
2005, Nucl. Phys. A, 752, 522

\bibitem[\protect\citeauthoryear{Crighton et al.}{2004}]{Crig2004}
Crighton N. H. M., Webb J. K., Ortiz-Gil A., Fernandez-Soto A.,
2004, MNRAS, 355, 1042

\bibitem[\protect\citeauthoryear{Cyburt et al.}{2008}]{Cyburt08}
Cyburt R.H., Fields B.D., Olive K.A., 2008, JCAP, 11, 12

\bibitem[\protect\citeauthoryear{Fields \& Olive}{2006}]{Fields2006}
Fields B. D., Olive K. A., 2006, Nucl. Phys. A, 777, 208

\bibitem[\protect\citeauthoryear{Foreman et al.}{2009}]{Foreman2009}
Foreman G., Volonteri M., Dotti M., 2009, ApJ, 693, 1554

\bibitem[\protect\citeauthoryear{Ivanov et al.}{2008}]{Iv2008}
Ivanov T. I., Roudjane M., Vieitez M. O., de Lange C. A.,
Tchang-Brillet W.-U L., Ubachs W., 2008, PRL, 100, 093007

\bibitem[\protect\citeauthoryear{Jenkins \& Peimbert}{1997}]{Jenkins97}
Jenkins E.B., Peimbert A., 1997, ApJ, 477, 265

\bibitem[\protect\citeauthoryear{Kirkman et al.}{2003}]{Kirk03}
Kirkman D., Tytler D., Suzuki N., O'Meara J.M., Lubin D., 2003,
ApJSS, 149, 1

\bibitem[\protect\citeauthoryear{Komatsu et al.}{2009}]{komatsu09}
Komatsu E., et al., 2009, ApJS, 180(2), 330

\bibitem[\protect\citeauthoryear{Lacour et al.}{2005a}]{Lacour05a}
Lacour S., et al., 2005a, A\&A, 430, 967

\bibitem[\protect\citeauthoryear{Lacour et al.}{2005b}]{Lacour05b}
Lacour S., et al., 2005b, ApJ, 627, 251

\bibitem[\protect\citeauthoryear{Le Bourlot}{2000}]{LeB2000}
Le Bourlot J., 2000, A\&A, 360, 656

\bibitem[\protect\citeauthoryear{Ledoux et al.}{2003}]{Ledoux03}
Ledoux C., Petitjean P., Srianand R., 2003, MNRAS, 346, 209

\bibitem[\protect\citeauthoryear{Le Petit et al.}{2002}]{LePetit02}
Le Petit F., Roueff E., Le Bourlot J., 2002, A\&A, 390, 369

\bibitem[\protect\citeauthoryear{Lepp et al.}{2002}]{Lepp02}
Lepp S., Stancil P.C., Dalgarno A., 2002, J. Phys. B, 35, R57

\bibitem[\protect\citeauthoryear{Levshakov \& Varshalovich}{1985}]{LV85}
Levshakov S. A., Varshalovich D.A., 1985, MNRAS, 212, 517

\bibitem[\protect\citeauthoryear{Levshakov et al.}{2002}]{LeV2002}
Levshakov S. A., Dessauges-Zavadsky M., D'Odorico S., Molaro P.,
2002, ApJ, 565, 696

\bibitem[\protect\citeauthoryear{Linsky et al.}{2006}]{Linsky06}
Linsky J. L., et al., 2006, ApJ, 647, 1106

\bibitem[\protect\citeauthoryear{McGreer \& Bryan}{2008}]{McGreer08}
McGreer I.D., Bryan G.L., 2008, ApJ, 685, 8

\bibitem[\protect\citeauthoryear{Noterdaeme et al.}{2007}]{Noterd2007}
Noterdaeme P., Ledoux C., Petitjean P., Le Petit F., Srianand R., Smette A., 2007, A\&A, 474, 393

\bibitem[\protect\citeauthoryear{Noterdaeme et al.}{2008a}]{Noterd2008a}
Noterdaeme P., Ledoux C., Petitjean P., Srianand R., 2008a, A\&A,
481, 327

\bibitem[\protect\citeauthoryear{Noterdaeme et al.}{2008b}]{Noterd2008b}
Noterdaeme P., Petitjean P., Ledoux C., Srianand R., Ivanchik
A., 2008b, A\&A, 491, 397

\bibitem[\protect\citeauthoryear{Olive et al.}{2000}]{Olive00}
Olive K.A., Steigman G., Walker T.P., 2000, Phys. Rep., 333, 389

\bibitem[\protect\citeauthoryear{O'Meara et al.}{2001}]{Omeara01}
O'Meara J. M., Tytler D., Kirkman D., Suzuki N., Prochaska J. X.,
Lubin D., Wolfe A. M., 2001, ApJ, 552, 718

\bibitem[\protect\citeauthoryear{O'Meara et al.}{2006}]{Omeara06}
O'Meara J. M., Burles S., Prochaska J. X., Prochter G. E., Bernstein
R. A., Burgess K. M., 2006, ApJ, 649, L61

\bibitem[\protect\citeauthoryear{Palla \& Galli}{1995}]{Palla95}
Palla F., Galli D., 1995, ApJ, 451, 44

\bibitem[\protect\citeauthoryear{Petitjean et al.}{2000}]{Petitj2000}
Petitjean P., Srianand R., and Ledoux C., 2000, A\&A, 364, L26

\bibitem[\protect\citeauthoryear{Pettini \& Bowen}{2001}]{Pettini01}
Pettini M., Bowen D. V., 2001, ApJ, 560, 41

\bibitem[\protect\citeauthoryear{Pettini et al.}{2008}]{Pettini08}
Pettini M., Zych B. J., Murphy M. T., Lewis A., Steidel~C.C., 2008,
MNRAS, 391, 1499

\bibitem[\protect\citeauthoryear{Puy et al.}{1993}]{Puy93}
Puy D., Al\'ecian G., Le Bourlot J., L\'eorat J., Pineau des For\^ets
G., 1993, A\&A, 267, 337

\bibitem[\protect\citeauthoryear{Sarkar}{1996}]{Sarkar96}
Sarkar S., 1996, Rep. on Progress in Phys., 95, 1493


\bibitem[\protect\citeauthoryear{Snow et al.}{2008}]{Snow2008}
Snow T.P., Ross T.L., Destree J.D., Drosback M.M., Jensen A.G.,
Rachford B.L., Sonnentrucker P., Ferlet R., 2008, ApJ, 688, 1124

\bibitem[\protect\citeauthoryear{Spitzer \& Cochran}{1973}]{Spit73}
Spitzer L. J., Cochran W. D., 1973, ApJ, 186, 23

\bibitem[\protect\citeauthoryear{Srianand et al.}{2008}]{Srian2008}
Srianand R., Noterdaeme P., Ledoux C., Petitjean P., 2008, A\&A,
482, L39

\bibitem[\protect\citeauthoryear{Srianand et al.}{2000}]{Srian2000}
Srianand R.,  Petitjean P., Ledoux C., 2000, Nature, 408, 931

\bibitem[\protect\citeauthoryear{Srianand et al.}{2005}]{Srian2005}
Srianand R.,  Petitjean P., Ledoux C., Ferland G., Gargi S., 2005, MNRAS, 362, 549

\bibitem[\protect\citeauthoryear{Steigman}{2007}]{Steig2007}
Steigman G., 2007, Ann. Rev. Nucl. Part. Sci., 57, 463

\bibitem[\protect\citeauthoryear{Vanden Berk}{2001}]{Vandenberg2001}
Vanden Berk, D. E., et al., 2001, AJ, 122, 549

\bibitem[\protect\citeauthoryear{Varshalovich et al.}{2001}]{Varsh2001}
Varshalovich D., Ivanchik A., Petitjean P., Srianand R., Ledoux
C., 2001, Astronomy Letters, 27, 683


\end{thebibliography}
\end{document}